\documentclass[twocolumn]{aa}

\usepackage{graphicx}
\usepackage{txfonts}
\usepackage{color}

\begin{document}

\title{An analysis of electron distributions in galaxy clusters
by means of the flux ratio of iron lines FeXXV and XXVI}

\author{D. A. Prokhorov\inst{1,2}, F. Durret\inst{1}, V. Dogiel \inst{3} and S. Colafrancesco \inst{4,5,6}}

\offprints{D.A. Prokhorov \email{prokhoro@iap.fr}}

\institute{Institut d'Astrophysique de Paris, CNRS, UMR 7095,
Universit\'{e} Pierre et Marie Curie, 98bis Bd Arago, F-75014
Paris, France
            \and
            Moscow Institute of Physics and Technology,
            Institutskii lane, 141700 Moscow Region, Dolgoprudnii, Russia
            \and
            Lebedev Physical Institute, 117924 Moscow, Russia
            \and
            ASI Science Data Center, ASDC c/o ESRIN,
            via G. Galilei 00044 Frascati, Italy.
            \and
            ASI, Viale Liegi 26, Roma, Italy
            \and
            INAF - Osservatorio Astronomico di Roma,
            via Frascati 33, I-00040 Monteporzio, Italy.
            }

\date{Accepted . Received ; Draft printed: \today}

\authorrunning{Prokhorov et al.}

\titlerunning{An analysis of electron distributions}

\abstract
{}
{The interpretation of hard X-ray emission from galaxy clusters is
still ambiguous and different models proposed can be probed using
various observational methods. Here we explore a new method based
on Fe line observations.}
{Spectral line emissivities have usually been calculated for a
Maxwellian electron distribution. In this paper a generalized
approach to calculate the iron line flux for a modified Maxwellian
distribution is considered.}
{We have calculated the flux ratio of iron lines for the various
possible populations of electrons that have been proposed to
account for measurements of hard X-ray excess emission from the
clusters A2199 and Coma. We found that the influence of the
suprathermal electron population on the flux ratio is more
prominent in low temperature clusters (as Abell 2199) than in high
temperature clusters (as Coma).}
{}


\maketitle

\section{Introduction}

Observations with BeppoSAX have detected hard X-ray tails in the
X-ray spectra of some clusters as the Coma cluster (Fusco-Femiano
et al. 1999) and Abell 2199 (Kaastra et al. 1998, Kaastra et al.
1999). These tails, which have been fit by power-law spectra, are
in excess to the thermal bremsstrahlung X-ray emission from the
hot intracluster medium (ICM). The evidence and the nature of hard
tails in these and other clusters is discussed in the recent
review by Rephaeli et al. (2008).
The hard X-ray fluxes from galaxy clusters are usually interpreted
either as due to inverse Compton scattering (ICS) of relativistic
electrons on relic photons (Sarazin \& Lieu 1998) or as
bremsstrahlung emission from nonthermal subrelativistic electrons
(see e.g. Sarazin \& Kempner 2000) or from thermal electrons with
a Maxwellian spectrum distorted by the particle acceleration
mechanism (Dogiel 2000, Dogiel et al. 2007).

The more traditional interpretation based on the ICS emission from
a relativistic electron population faces a serious problem. The
combination of hard X-ray and radio observations of the Coma
cluster within the ICS model strongly indicates a very low
magnetic field, B$\sim$0.1 $\mu$G, much lower than the values
derived from Faraday rotation measurements (see e.g. Clarke et al.
2001). The situation in Abell 2199 is more extreme because no
extended diffuse radio emission is detected from this cluster. The
discovery of the hard X-ray emission of the cluster A2199 implies
a very weak ICM magnetic field of $\lesssim 0.07\mu$G if the hard
X-ray emission is ICS (Kempner \& Sarazin 2000).

Bremsstrahlung radiation from suprathermal electrons with energies
higher than 10 keV (nonthermal electrons or thermal electrons with
a distorted Maxwellian spectrum)  may explain the hard X-ray
excess emission observed in the Coma cluster and Abell 2199 (e.g.
Sarazin \& Kempner 2000, Dogiel 2000). This is an alternative to
the traditional but problematic inverse Compton scattering
interpretation. Such subrelativistic electrons would form a
particle population in excess over the thermal gas. A possible
explanation for this population would be that they are particles
being accelerated to higher energies, either by intracluster
shocks or by turbulence in the ICM (e.g. Dogiel 2000).

The best way to resolve the question of whether the observed hard
X-rays are due to ICS or are evidence for a modified thermal
distribution in clusters is to probe directly such a distribution.

The Sunyaev-Zel'dovich (SZ) effect signal, the spectrum of which
depends on the electron distribution function in clusters of
galaxies, can be used to discriminate among different
interpretations of the X-ray excess (Colafrancesco 2007, Dogiel et
al. 2007). The study of the influence of suprathermal electrons on
the SZ effect was done for the Coma and Abell 2199 clusters by
Blasi et al. (2000) and Shimon \& Rephaeli (2002). However,
realistical models of suprathermal electrons in Coma and Abell
2199 predict a spectral distortion of the cosmic microwave
background radiation due to these electrons that is only a small
fraction of the corresponding SZ effect due to the hot
intracluster gas (see e.g. Shimon \& Rephaeli 2002, Dogiel et al.
2007). Therefore the observation of the impact of suprathermal
electrons on the cosmic microwave background will be challenging
from the experimental side (see Dogiel et al. 2007).

In this paper we consider a new probe to discriminate among
different interpretations of the X-ray excess, namely the flux
ratio of the emission lines due to FeK$\alpha$ transitions: FeXXV
(helium-like) and FeXXVI (hydrogen-like). This flux ratio is very
sensitive to the population of electrons with energies higher than
the ionization potential of a FeXXV ion (which is $\approx$ 8.8
keV) and is a promising tool to reveal the presence of
suprathermal electrons in galaxy clusters.

A generalized approach to calculate the iron line fluxes is
considered in Sect. 2. The flux ratio of the emission lines due to
the FeK$\alpha$ transitions is calculated for the modified thermal
distributions in the clusters Abell 2199 and Coma in Sect. 3. The
possibility to separate the thermal and non-thermal components by
using the shape of the bremsstrahlung continuum spectrum is
discussed in Sect. 4. We draw our conclusions in the final Sect.
5.

\section{The flux ratio of the FeXXV and XXVI iron lines.}

Since the fluxes of the FeXXV and FeXXVI lines have the same
dependence on the metal abundance, as well as on the emission
measure, their ratio is independent of these parameters. This iron
line ratio can be used to determine the temperature of the
intracluster gas (e.g. Nevalainen et al. 2003). In this section we
propose a generalized approach to calculate the iron line flux
ratio for modified Maxwellian electron distributions.

\subsection{Ionization and recombination rates.}

The ionization rates, recombination rates and emissivity in a
spectral line have usually been calculated for a Maxwellian
electron distribution (e.g. Arnaud \& Raymond 1992). However, in
many low-density astrophysical plasmas, the electron distribution
may differ from a Maxwellian distribution. The influence of the
shape of the electron distribution on the ionization and
recombination rates in various physical conditions was examined by
Porquet et al. (2001).

A Maxwellian distribution is generally considered for the electron
distribution in galaxy clusters. Modified Maxwellian electron
distributions expected in galaxy clusters with a hard X-ray excess
seem to be reasonably described by a Maxwellian distribution at
low energy and by a power-law distribution at higher energy (e.g.
Sarazin \& Kempner 2000).

It is convenient to express the electron distribution in term of
the reduced energy $x=E/kT$:
\begin{eqnarray}
dn_{\mathrm{e}}(x)=n_{\mathrm{e}} f(x) dx \; ,
\end{eqnarray}
where $n_{\mathrm{e}}$ is the electronic density and $k$ is the Boltzmann
constant.

Let us consider a collisional process with cross section $\sigma(E)$, varying
with the energy $E$ of the incident electron. The corresponding rate
coefficient $\Gamma$ (cm$^3$s$^{-1}$) either for a Maxwellian distribution or a
modified thermal distribution, $f(x)$, is obtained by averaging the product of
the cross section by the electron velocity over the electron distribution
function:

\begin{equation}
\Gamma=\left(\frac{2kT}{m_{\mathrm{e}}}\right)^{1/2}
\int^{\infty}_{x_{\mathrm{thr}}} x^{1/2} \sigma(xkT) f(x) dx
\end{equation}
where $m_{\mathrm{e}}$ is the electron mass,
$x_{\mathrm{thr}}=E_{\mathrm{thr}}/(kT)$, $E_{\mathrm{thr}}$
corresponds to the threshold energy of the considered process.

For recombination processes, no threshold energy is involved and
$x_{\mathrm{thr}}=0$. The rates are noted $C_{\mathrm{I}}$,
$\alpha_{\mathrm{RR}}$ and $\alpha_{\mathrm{DR}}$ for the
ionization, radiative and dielectronic recombination processes
respectively, for a Maxwellian electron distribution.

In equilibrium, the ionic fractions do not depend on the electron
density and the ionic fraction ratio
$\xi_{\mathrm{FeXXV}}/\xi_{\mathrm{FeXXVI}}$ of two adjacent
stages FeXXV and FeXXVI for a Maxwellian distribution can be
expressed by:

\begin{equation}
\left(\frac{\xi_{\mathrm{FeXXV}}}{\xi_{\mathrm{FeXXVI}}}\right)_{\mathrm{M}}=
\frac{\alpha_{\mathrm{R}}(\mathrm{FeXXVI})}{C_{\mathrm{I}}(\mathrm{FeXXV})}
\end{equation}
where $C_{\mathrm{I}}(\mathrm{FeXXV})$ and
$\alpha_{\mathrm{R}}(\mathrm{FeXXVI})$ are the ionization and
total recombination rates of ions FeXXV and FeXXVI respectively.

For the direct ionization cross section of FeXXV we use the
parametric formula as in Arnaud \& Rothenflug (1985):
\begin{equation}
\sigma_{\mathrm{DI}}(E)=\sum_{\mathrm{j}}
\frac{A_{\mathrm{j}}U_{\mathrm{j}}+B_{\mathrm{j}}U^{2}_{\mathrm{j}}+
C_{\mathrm{j}}\ln(u_{\mathrm{j}})+D_{\mathrm{j}}\ln(u_{\mathrm{j}})/u_{\mathrm{j}}}
{u_{\mathrm{j}}I^2_{\mathrm{j}}}
\end{equation}
with $u=E/I_{\mathrm{j}}$, $U_{\mathrm{j}}=1-1/u_{\mathrm{j}}$,
$E$ is the incident electron energy, $I_{\mathrm{j}}$ is the
collisional ionization potential for the level j considered.

The sum is performed over the subshells j of the ionized ion, for
the ion FeXXV the 1s subshell is considered (Arnaud \& Raymond
1992). The parameters $A$, $B$, $C$, $D$ (in units of $10^{-14}$
cm$^2$ eV$^2$) and $I$ (in eV) are taken from Arnaud \& Raymond
(1992).

The autoionization process of the FeXXV ion can be neglected
(Arnaud \& Raymond 1992).

The ratio of the ionization rate in a modified Maxwellian
distribution over that in a Maxwellian distribution is:
\begin{equation}
\beta_{\mathrm{I}}=\frac{\int^{\infty}_{x_{\mathrm{thr}}}
x^{1/2}\sigma_{\mathrm{DI}}(xkT) f_{\mathrm{MM}}(x)
dx}{\int^{\infty}_{x_{\mathrm{thr}}}
x^{1/2}\sigma_{\mathrm{DI}}(xkT) f_{\mathrm{M}}(x) dx}
\end{equation}
where $f_{\mathrm{MM}}(x)$ is the modified Maxwellian
distribution, $f_{\mathrm{M}}(x)$ is the Maxwellian distribution,
$\sigma_{\mathrm{DI}}$ is the direct ionization cross section of
the ion FeXXV. The ionization rate is very sensitive to the
fraction of electrons above the threshold energy.

Recombination of a free electron can proceed either through a
radiative free-bound transition $(\mathrm{FeXXVI}+e^{-}\rightarrow
\mathrm{FeXXV}+h\nu)$ or by a radiationless dielectronic
recombination.

The radiative recombination rates are less affected by a modified
thermal distribution than the ionization rates, since the cross
section for recombination decreases with energy and there is no
threshold. To estimate the radiative recombination rate ratio, we
follow the method used by Owocki \& Scudder (1983) and Porquet et
al. (2001). The ratio of the radiative recombination rate in a
modified Maxwellian distribution over that in a Maxwellian
distribution is:
\begin{equation}
\beta_{\mathrm{RR}}=\frac{\int^{\infty}_{0} x^{-\eta}
f_{\mathrm{MM}}(x) dx}{\int^{\infty}_{0} x^{-\eta}
f_{\mathrm{M}}(x) dx}. \label{betarr}
\end{equation}
Following the method of Porquet et al. (2001) we use the value
$\eta=0.8$ for an iron ion corresponding to the mean value
$<\eta>$ reported in Arnaud \& Rothenflug (1985).

The dielectronic recombination is a resonant process involving
bound states at discrete energies $E_{i}$ and can be computed by
summing the contribution of many such bound states. Following the
method used by Owocki \& Scudder (1983), we assume that the
corresponding dielectronic recombination cross section can be
approximated by:
\begin{equation}
\sigma_{DR}(E)=\sum_{\mathrm{i}} D_{\mathrm{i}}
\delta(E-E_{\mathrm{i}})
\end{equation}
where $D_{\mathrm{}i}$ are the dielectronic recombination
coefficients.

The ratio of the dielectronic recombination rate in a modified
Maxwellian distribution over that in a Maxwellian distribution is:
\begin{equation}
\beta_{\mathrm{DR}}=\frac{\int^{\infty}_{0}
x^{1/2}\sigma_{\mathrm{DR}}(xkT) f_{\mathrm{MM}}(x)
dx}{\int^{\infty}_{0} x^{1/2}\sigma_{\mathrm{DR}}(xkT)
f_{\mathrm{M}}(x) dx}.
\end{equation}

For the ion FeXXVI there is one bound state with the energy
$E_{1}=5.3$ keV. If the break energy up to which the electron
distribution is Maxwellian is higher than this bound state energy
then the dielectronic recombination rate is not influenced by the
suprathermal electrons.

Thus the ionic fraction ratio
$\xi_{\mathrm{FeXXV}}/\xi_{\mathrm{FeXXVI}}$ of two adjacent
stages FeXXV and FeXXVI for a modified Maxwellian distribution can
be written as:

\begin{equation}
\left(\frac{\xi_{\mathrm{FeXXV}}}{\xi_{\mathrm{FeXXVI}}}\right)_{\mathrm{MM}}=
\frac{\beta_{\mathrm{RR}}\alpha_{\mathrm{RR}}(\mathrm{FeXXVI})+
\beta_{\mathrm{DR}}\alpha_{\mathrm{DR}}(\mathrm{FeXXVI})}{\beta_{\mathrm{I}
}C_{\mathrm{I}}(\mathrm{FeXXV})}.
\end{equation}

\subsection{Excitation rates and iron line flux ratio.}

In the coronal model (see e.g. Mewe 1999), the line spectrum is
dominated by radiative decay following electron impact excitation,
plus a smaller contribution of recombination lines. We assume here
that all iron ions that are to be excited are in the ground state
(see e.g. Mewe \& Gronenschild 1981). Considering only the
dominant process of collisional excitation (Tatischeff 2003), the
volume emissivity $P^{ab}_{\mathrm{Fe}^{+i}}$ (in units of photons
cm$^{-3}$ s$^{-1}$) of a particular line transition $a\rightarrow
b$ in an ion Fe$^{+\mathrm{i}}$ can be written as
\begin{equation}
P^{ab}(\mathrm{Fe}^{+i})=n_{\mathrm{e}} n_{\mathrm{H}} a_{\mathrm{Fe}}
\xi_{\mathrm{Fe}^{+i}} S^{ga}_{\mathrm{Fe}^{+i}} B_{ab} \; ,
 \label{pab}
\end{equation}
where $n_{H}$ is the H ionic number density (cm$^{-3}$), $a_{\mathrm{Fe}}$ is
the abundance of iron relative to hydrogen, $\xi_{\mathrm{Fe}^{+i}}$ is the
ionic fraction of ion Fe$^{+i}$, $S^{ga}_{\mathrm{Fe}^{+i}}$ is the rate for
electron impact excitation of an ion Fe$^{+i}$ from its ground state to its
excited state a, and $B_{ab}$ is the radiative branching ratio of the
transition $a \rightarrow b$ among all possible transitions from the level $a$.

The excitation rates $S^{ga}_{\mathrm{Fe}^{+i}}[f(x)]$ are
functionals which are calculated by averaging the product of the
corresponding cross section and electron velocity over electron
distribution functions $f(x)$:

\begin{equation}
S^{ga}_{\mathrm{Fe}^{+i}}[f(x)] =
\left(\frac{2kT}{m_{\mathrm{e}}}\right)^{1/2}\int^{\infty}_{x_{\mathrm{thr}, \,
\mathrm{ex}}} x^{1/2}\sigma^{\mathrm{Fe}^{+i}}_{\mathrm{ex}}(x k T) f(x) dx \;
,
\end{equation}
where $\sigma^{\mathrm{Fe}^{+i}}_{\mathrm{ex}}$ is the excitation section of
the ion Fe$^{+i}$, $x_{\mathrm{thr},\, \mathrm{ex}} =E_{\mathrm{thr},\,
\mathrm{ex}}/(kT)$, and $E_{\mathrm{thr},\, \mathrm{ex}}$ corresponds to the
threshold energy of the excitation process.

The emission lines due to FeK$\alpha$ transitions of ions FeXXV
and FeXXVI are at 6.7 keV and 6.9 keV respectively. Ions with
closed-shell configurations are more stable than those with
partially filled shells; thus He-like FeXXV, whose ground state is
$1s^2$, is dominant in a large temperature range, because its
ionization rate is relatively low compared to those of adjacent
ions (e.g. Arnaud \& Raymond 1992). The strongest line emission is
then the He-like FeK$\alpha$  line complex at 6.7 keV that
corresponds to transitions $1s^2-1s2p ^{1}P$, $1s^2-1s2p^{3}P$,
$1s^2-1s2s^{3}S$. At high temperatures (e.g., $kT=8$ keV), the
hydrogen-like iron line ($2p \rightarrow 1s$ transition) at 6.9
keV also becomes intense.

The electron impact excitation scaled cross sections for the
helium-like ion FeXXV (including impact excitation from $1^{1}S$
to $2^{1}P$, $2^{3}P$, $2^{3}S$ levels) are taken from the article
of Bazylev \& Chibisov (1981):

\begin{equation}
Z^4 \sigma(1s^2\rightarrow
1s2p^1P)=\left(\frac{1.93}{z^2}+\frac{6.07 \ln(z)}{z}\right) \pi
a^2_{0}
\end{equation}

\begin{equation}
Z^4 \sigma(1s^2\rightarrow 1s2p^3P)=\frac{2.04}{z} \pi a^2_{0}
\end{equation}

\begin{equation}
Z^4 \sigma(1s^2\rightarrow 1s2s^3S)=
\left(\frac{0.93}{z^3}-\frac{0.59}{z^4}\right) \pi a^2_{0}
\end{equation}
where $a_{0}$ is the Bohr radius, $z$ is the incident electron
energy in threshold units, $Z$ is the ion nuclear charge.

The electron impact excitation scaled cross section for the
hydrogen-like ion FeXXVI is taken from the paper of Fisher et al.
(1997):

\begin{equation}
Z^4 \sigma(1s\rightarrow 2p)=
\left(\frac{1.66}{z^2}+2.49\frac{\ln(z)}{z}\right) \pi a^2_{0}.
\end{equation}

There are also contributions from excitations to higher levels,
that may radiatively decay to the upper levels of the He-like
triplet and H-like doublet. These so-called cascade effects
generally cannot be ignored. The electron impact excitation scaled
cross sections of an iron ion from its ground state to the higher
levels are taken from the article of Bazylev \& Chibisov (1981).

The flux ratio of iron lines FeXXV and FeXXVI taking into account
electron impact excitation is then

\begin{equation}
R_{\mathrm{ei}}=\frac{P^{1-2}(\mathrm{FeXXV}) \Delta
E^{1-2}_{\mathrm{FeXXV}}}{P^{1-2}(\mathrm{FeXXVI}) \Delta
E^{1-2}_{\mathrm{FeXXVI}}} \; ,
\end{equation}
where the volume emissivities $P^{1-2}(\mathrm{FeXXV})$ and
$P^{1-2}(\mathrm{FeXXVI})$ are for the He-like triplet and for the
H-like doublet, and the energies $\Delta E^{1-2}_{\mathrm{FeXXV}}$
and $\Delta E^{1-2}_{\mathrm{FeXXVI}}$ equal 6.7 keV and 6.9 keV
respectively.

According to Eq. (\ref{pab}) the expression for the flux ratio in
terms of ionic fractions and excitation rates writes as
\begin{equation}
R_{{\mathrm{ei}}}=\frac{\Delta E^{1-2}_{\mathrm{FeXXV}}}{\Delta
E^{1-2}_{\mathrm{FeXXVI}}}\frac{\xi_{\mathrm{FeXXV}}} {\xi_{\mathrm{FeXXVI}}
}\frac{Q^{1-2}_{\mathrm{FeXXV}}}{Q^{1-2}_{\mathrm{FeXXVI}} } \; ,
\end{equation}
where the rate coefficients are
$Q^{1-2}_{\mathrm{FeXXV}}=\sum\limits_{a}\sum\limits_{b (<a)}
S^{1s^2-a}_{\mathrm{FeXXV}} B_{ab}$,
 $Q^{1-2}_{\mathrm{FeXXVI}}=\sum\limits_{a}\sum\limits_{b
(<a)} S^{1s-a}_{\mathrm{FeXXVI}} B_{ab}$.  The excited states $b$
correspond to the upper levels of the He-like triplet and of the
H-like doublet, and the radiative branching ratios are given by
\begin{equation}
B_{ab}=\frac{A_{ab}}{\sum\limits_{c (<a)} A_{ac}} \; .
\end{equation}
All necessary transition probabilities $A_{ac}$ are taken from
Ralchenko et al. (2008).

However, in plasmas in collisional ionization equilibrium,
radiative recombination still contributes about 10\% to the total
line flux. We calculated the rate coefficients for the
contribution from radiative recombination to the spectral line
formation with equation (A.9) in Mewe et al. (1985). The influence
of the suprathermal electron population on the radiative
recombination rates is described by Owocki \& Scudder (1983) (see
Eq. \ref{betarr}). Although, as noted below, the ratio of the
radiative recombination rate $\beta_{RR}$ depends slightly on the
presence of suprathermal electrons in the spectrum, the ionic
fractions of FeXXVI and of FeXXVII and, therefore, the line
emissivities change with the presence of suprathermal electrons.

The line flux ratio taking into account both electron impact
excitation and radiative recombination is given by
\begin{equation}
R=\frac{\Delta E^{1-2}_{\mathrm{FeXXV}}}{\Delta
E^{1-2}_{\mathrm{FeXXVI}}}\times\frac{\xi_{\mathrm{FeXXV}}Q^{1-2}_{\mathrm{FeXXV}}
+ \xi_{\mathrm{FeXXVI}}\alpha^{1-2}_{\mathrm{RR},\
\mathrm{FeXXV}}} {\xi_{\mathrm{FeXXVI}}Q^{1-2}_{\mathrm{FeXXVI}} +
\xi_{\mathrm{FeXXVII}}\alpha^{1-2}_{\mathrm{RR},\ \mathrm{FeXXVI}}
; ,}
\end{equation}
where $\alpha^{1-2}_{\mathrm{RR},\ \mathrm{FeXXV}}$ and
$\alpha^{1-2}_{\mathrm{RR},\ \mathrm{FeXXVI}}$ are the rate
coefficients for the contribution from radiative recombination to
the spectral lines FeXXV (He-like triplet) and FeXXVI (H-like
doublet) respectively. The ionic fractions of
$\xi_{\mathrm{FeXXVII}}$ and $\xi_{\mathrm{FeXXVI}}$ were
calculated following the same method as in Sect. 2.1.

Figure 1 shows the iron line flux ratios for the pure Maxwellian
distribution calculated in this section (dot-dashed line) and
those obtained from the MEKAL model (points) by Nevalainen et al.
(2003).

\section{An analysis of electron spectra in clusters.}

Bremsstrahlung from suprathermal electrons has been invoked as a
possible explanation for hard X-ray tails in the X-ray spectra of
some galaxy clusters. In this section we calculate the impact of
suprathermal electrons on the FeXXV and FeXXVI emission line flux
ratio in the clusters A2199 and Coma.

\subsection{The galaxy cluster Abell 2199}

A2199 is a bright cluster at redshift z=0.03. Its average gas
temperature is $kT=4.7$ keV (Kaastra et al. 1999). Spatially
resolved spectroscopy shows a hard tail in the X-ray spectrum of
this galaxy cluster (Kaastra et al. 1998). To interpret this hard
tail Sarazin \& Kempner (2000) applied a non-thermal
bremsstrahlung model with the electron distribution function
$f^{(1)}_{MM}(x)$ given by:
\begin{eqnarray}
f^{(1)}_{\mathrm{MM}}(x)= & f_{\mathrm{M}}(x), & \ \ \ \ x<3  \nonumber \\
f^{(1)}_{\mathrm{MM}}(x)=& f_{\mathrm{M}}(x) + \lambda
x^{-(\mu+1)/2}, & \ \ \ \ x\geqslant3
\end{eqnarray}
where $\mu=3.33$, $\lambda=0.34$ is found from the condition that
the non-thermal electron population is 8.1\% of the thermal
population (Sarazin \& Kempner 2000).

The ratio of the radiative recombination rates $\beta_{\mathrm{RR}}$ (see Eq.
\ref{betarr}) is 1.013 for the electron distribution function
$f^{(1)}_{\mathrm{MM}}(x)$. Since the break energy of $3kT$ is higher than the
bound state energy 5.3 keV for the cluster temperature, the ratio of the
dielectronic recombination rates is 1. Therefore the recombination rates are
not affected by suprathermal electrons.

In Fig. 1 we compare the flux ratios $R$ for a Maxwellian electron distribution
(dot-dashed line) and for a modified Maxwellian distribution
$f^{(1)}_{\mathrm{MM}}$ (dashed line) in the temperature range 4.5 keV $<$ kT
$<$ 8.5 keV.

For the galaxy cluster A2199 (kT=4.7 keV) the flux ratio R for a
modified Maxwellian distribution $f^{(1)}_{\mathrm{MM}}$ decreases
by $\approx27\%$ with respect to the case of a Maxwellian
distribution. This value of the flux ratio would correspond to a
thermal electron spectrum (i.e. without suprathermal electrons)
with an effective temperature of kT = 5.4 keV.

\subsection{The Coma cluster.}

The Coma cluster is a rich, hot, nearby (z=0.02) galaxy cluster.
Its average temperature is $kT=8.2$ keV as derived from XMM-Newton
observations (Arnaud et al. 2001).

Hard X-ray radiation was detected in excess of thermal emission in the Coma
cluster by a first Beppo-SAX observation (Fusco-Femiano et al. 1999) and
confirmed by a second independent observation with a time interval of about 3
yr (Fusco-Femiano et al. 2004). The reliability of the Fusco-Femiano et al.
(1999, 2004) analyses was further discussed by Rossetti \& Molendi (2004, 2007)
and by Fusco-Femiano et al. (2007).

The presence of a second component in the X-ray spectrum of the
Coma cluster has also been derived from two RXTE observations
(Rephaeli \& Gruber 2002).

The spectrum of background and accelerated electrons was found by Gurevich
(1960) from a kinetic equation describing stochastic particle acceleration:
\begin{eqnarray}
f^{(2)}_{\mathrm{MM}}(x)= \frac{2\sqrt{x}}{\sqrt{\pi}}
\left(\exp\left(-\int^{\sqrt{2x}}_{0}\frac{g dg}{1+\alpha
g^5}\right)-\exp\left(-\int^{\infty}_{0}\frac{g dg}{1+\alpha
g^5}\right)\right)\nonumber \\
& &\label{f2}
\end{eqnarray}
where the parameter $\alpha=9\cdot10^{-4}$ was derived by Dogiel
(2000) from the bremsstrahlung model for the origin of the hard
X-ray emission from the Coma cluster.
Here $x=E/(kT)$ is the reduced energy and the integration has been
done on the quantity $g=p/\sqrt{m_{\mathrm{e}}kT}$ that is the
dimensionless momentum.

\begin{figure}[ht]
\centering
\includegraphics[angle=0, width=8cm]{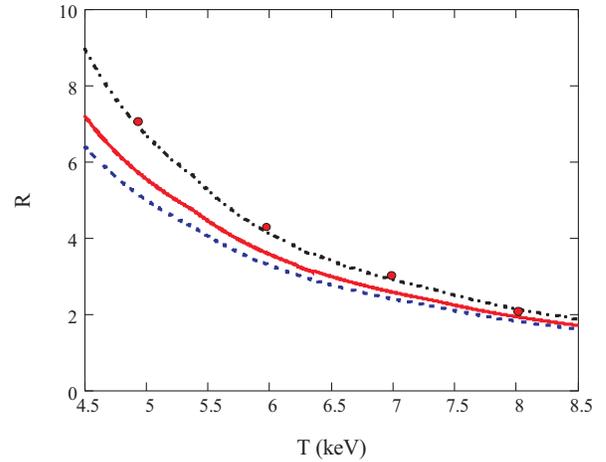}
\caption{Iron line flux ratios $R$ for a Maxwellian electron
distribution (dot-dashed line), a modified Maxwellian distribution
$f^{(1)}_{\mathrm{MM}}$ (dashed line) and a modified Maxwellian
distribution $f^{(2)}_{\mathrm{MM}}$ (solid line) in the
temperature range 4.5 keV $<$ kT $<$ 8.5 keV}
\end{figure}

The ratio of the radiative recombination rates $\beta_{\mathrm{RR}}$ (see Eq.
\ref{betarr}) is 1.01 for the electron distribution function
$f^{(2)}_{\mathrm{MM}}(x)$. For values $kT$ in the range $4.5-8.5$ keV, the
values of $\beta_{\mathrm{DR}}$ are found in the range 1.0005-1.0025. Therefore
the recombination rates are not affected by suprathermal electrons.

In Fig. 1 we compare the flux ratios R for the Maxwellian electron
distribution (dot-dashed line) and the modified Maxwellian
distribution $f^{(2)}_{\mathrm{MM}}$ (solid line) in the
temperature range 4.5 keV $<kT<$ 8.5 keV.

For the Coma cluster ($kT$=8.2 keV) the flux ratio $R$ for a
modified Maxwellian distribution $f^{(2)}_{\mathrm{MM}}$ decreases
by $\approx9\%$ with respect to the case of a Maxwellian
distribution. This value of the flux ratio would correspond to a
thermal electron spectrum (i.e. without suprathermal electrons)
with the effective temperature of kT = 8.6 keV.

\subsection{A synthetic low temperature cluster}

In Sects. 3.1 and 3.2 iron line flux ratios were calculated for
low temperature and high temperature galaxy clusters (Abell 2199
and Coma respectively). As shown in Fig.1, the impact of a
suprathermal electron population on the iron line flux ratio is
stronger in low temperature clusters. We demonstrate here for a
specific example how the effective temperature inferred from the
flux ratio of the iron lines can yield important constraints on
the fraction of suprathermal electrons. For this purpose, a
synthetic cluster with temperature $kT=4.7$ keV and with an
electron distribution function $f^{(1)}_{\mathrm{MM}}$ is
considered. The dependence of the effective temperature on the
fraction of suprathermal electrons is shown in Fig. 2.

\begin{figure}[ht]
\centering
\includegraphics[angle=0, width=8cm]{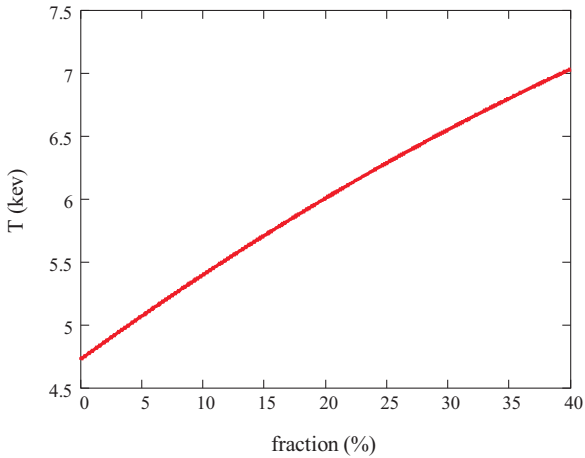}
\caption{Dependence of the effective temperature $kT$ (given in units of keV)
on the fraction of suprathermal electrons in a cluster of temperature 4.7 keV.}
\end{figure}

Since the effective temperature changes strongly with the
suprathermal electron fraction, the iron flux ratio can be used to
reveal a suprathermal electron population in low temperature
clusters and in cluster cool cores.

\section{The continuum spectrum}

Non-thermal electrons change the flux ratio R and lead to an
apparently higher temperature derived from the iron line ratio
(i.e. the effective temperature, see Sect. 3 and Fig. 2).
Alternatively, the temperature can be measured from the
bremsstrahlung spectrum curvature. But, in the presence of
non-thermal electrons, these same electrons will also give rise to
non-thermal bremsstrahlung, therefore altering the shape of the
continuum spectrum. A method to determine the temperature of the
thermal part of a more complex electron distribution from the
bremsstrahlung spectrum is proposed in this section. A
disagreement between both temperatures will depend on the strength
of the non-thermal electron component.

To separate the contributions of thermal (low-energy) and of
non-thermal (high-energy) electron components to the
bremsstrahlung spectrum we study here the features of the energy
flux spectrum.

The bremsstrahlung energy flux can be estimated as
\begin{equation}
\Phi(E_{\mathrm{x}})= \frac{n^2_{\mathrm{e}} V}{4\pi d^2}
E_{\mathrm{x}}\int^{\infty}_{E_{\mathrm{x}}}
\sigma_{\mathrm{B}}(E, E_{\mathrm{x}})
\sqrt{\frac{2E}{m_{\mathrm{e}}}} f(E) dE\label{phi}
\end{equation}
where V is the cluster volume, d is the distance to the galaxy
cluster, and $\sigma_{\mathrm{B}}(E, E_{\mathrm{x}})$ is the
bremsstrahlung cross section.
For the sake of illustration, let us consider the bremsstrahlung
cross section in the form of $\sigma_{\mathrm{B}}(E,
E_{\mathrm{x}})=const/E_{\mathrm{x}}E$ (following the
approximation from the paper of Ginzburg 1979). Then, for a
Maxwellian electron distribution, the dependence of the energy
flux on the photon energy is given by $\Phi(E_{\mathrm{x}})\propto
\exp(-E_{\mathrm{x}}/kT)$. For modified Maxwellian electron
distributions, the energy flux at low photon energies has two
components: thermal $\Phi_{1}(E_{\mathrm{x}})=C_{1}
\exp(-E_{\mathrm{x}}/kT)$ and non-thermal
$\Phi_{2}(E_{\mathrm{x}})=C_{2}$. The non-thermal component of the
bremsstrahlung energy flux is constant if the low limit of the
integral in Eq.(\ref{phi}) is smaller than the energy
$E_{\mathrm{M}}$ at which the electron distribution deviates from
a Maxwellian distribution ($E_{\mathrm{M}}=14$ keV for the cluster
A2199 (Kempner et al. 2000) and $E_{\mathrm{M}}=30$ keV for the
Coma cluster (Dogiel 2000)). Considering the energy band with
$E_{\mathrm{x}}<E_{\mathrm{M}}$, it is possible to fit the energy
flux spectrum with a function $\Phi(E_{\mathrm{x}})=C_{1}
\exp(-E_{\mathrm{x}}/kT)+C_{2}$ and calculate the normalization
constants $C_{1}$ and $C_{2}$, and also the temperature $T$ of the
thermal part of the electron distribution from the bremsstrahlung
spectrum.

The analysis of RXTE measurements for the Coma cluster (Rephaeli
\& Gruber 2002) has shown evidence for the presence of a second
spectral component at energies up to $\sim$20 keV, since the fit
to a single isothermal model has a poor quality. When a second
thermal component is added, the best fit temperatures of the
primary and secondary components are then $kT_{1}=7.5$ keV and the
very high value $kT_{2}\simeq37.1$ keV. The contribution of the
second component $\Phi_{2}(E_{\mathrm{x}})\propto
\exp(-E_{\mathrm{x}}/kT_{2})$ to the low-energy continuum spectrum
is flat as noted above. Therefore the thermal and non-thermal
components can, in principle, be separated by studying the shape
of the continuum spectrum.

From Eq.(\ref{phi}) we calculated the energy fluxes
$\Phi(E_{\mathrm{x}})$ of the Coma cluster in the range 4-20 keV
for a Maxwellian electron distribution and for a modified
Maxwellian distribution $f^{(2)}_{\mathrm{MM}}$, as shown in Fig.
3. The values of the cluster parameters (temperature, density etc)
were taken from Dogiel (2000). The difference between the total
4-20 keV fluxes (i.e. the total 4-20 keV flux of the second
spectral component) is $\approx$7\%. The spectral components must
be separated in order to obtain the temperature from the
bremsstrahlung continuum spectrum.

\begin{figure}[ht]
\centering
\includegraphics[angle=0, width=8cm]{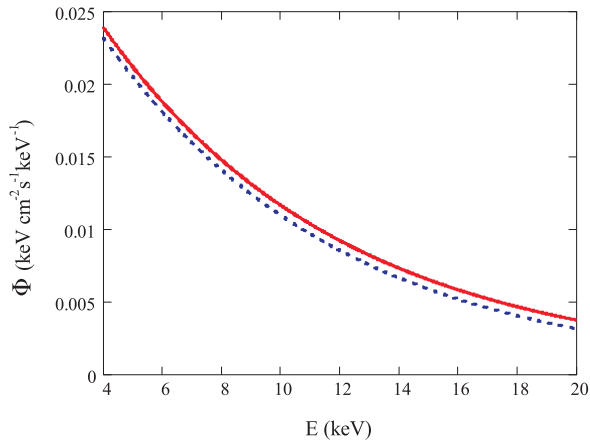}
\caption{The energy flux $\Phi(E_{\mathrm{x}})$ of the Coma
cluster in the range 4-20 keV  for a Maxwellian electron
distribution (dashed line) and for a modified Maxwellian
distribution $f^{(2)}_{\mathrm{MM}}$ (solid line).}
\end{figure}

\section{Conclusions}

We have shown in this paper that the iron line flux ratio depends
on the presence of suprathermal electrons that have been proposed
to account for measurements of hard X-ray excess emission from
galaxy clusters. The influence of the energetic suprathermal
electron population on the iron line flux ratio is more prominent
in low temperature clusters (as Abell 2199) than in high
temperature clusters (as Coma) because the fraction of thermal
electrons with energies higher than the helium-like iron
ionization potential in low temperature clusters is smaller than
that in high temperature clusters.

Since the decrease of the flux ratio of He-like $K_{\alpha}$ to
H-like $K_{\alpha}$ lines is expected for modified Maxwellian
distributions in A2199 and Coma with respect to the case of a
Maxwellian distribution, observation of the flux ratio is a tool
to test the nonthermal electron bremsstrahlung model and to
discriminate among different interpretations of the X-ray excess.
In order to demonstrate the presence and to measure the strength
of the non-thermal electron component we propose to compare the
temperatures obtained from the iron line flux ratio and from the
low-energy continuum spectrum.

The spectral resolution of XMM-Newton is sufficient to measure the
flux ratio of the iron lines in hot temperature clusters. The
constraint of the flux ratio for Coma within radius 5$^{\prime}$
is $1.6^{+0.9}_{-0.6}$ (Nevalainen et al. 2003). However, the
XMM-Newton sensitivity in this high-temperature regime is
insufficient to reveal the contribution from suprathermal
electrons in the Coma cluster (see Fig. 1).

At low temperatures (e.g. $kT<5$ keV) the FeXXVI line is weak and
is within the noise level of the XMM-Newton data (Nevalainen et
al. 2003). Therefore the iron line ratio cannot be measured by
XMM-Newton in cooler clusters. On the other hand, the flux ratio
of the iron lines in low temperature clusters can be detectable by
XMM-Newton if the fraction of suprathermal electrons is
sufficiently high (see Fig. 2).

Suzaku is also able to measure these two Fe lines in hot clusters
due to its good spectral resolution (e.g. Fujita et al. 2008).

We have considered the He-like triplet and the H-like doublet iron
lines in this paper. Although XMM-Newton and Suzaku can
distinguish the He-like from the H-like complex, their spectral
resolution of $\sim$100 eV causes that the observed line features
do not only consist of the pure He-like triplet and H-like
doublet, but each of these is blended with a multitude of
satellite lines (e.g. Gabriel 1972; Dubau et al. 1981). For
instance, for a temperature of 4.5 keV about $\sim$30\% of the
flux from both line complexes is due to these satellite lines. In
order to analyse the influence of satellite lines on the
measurement precision of the iron line flux ratio we calculated
the flux ratio $R_{B}$ of the two (6.6-6.7 keV) and (6.9-7.0 keV)
blends using the line list which was taken from
http://www.sron.nl/ divisions/hea/spex/version1.10/line/index.html
and found that in the temperature range [4.5-8.5 keV] the values
of R and $\mathrm{R}_{\mathrm{B}}$ differ by less than 5\%.
Therefore the blend flux ratio $\mathrm{R}_{\mathrm{B}}$ as well
as the ratio R can be used to measure the temperature in this
temperature range.

The autoionizing levels responsible for the satellites are excited
by electrons at precisely the energies $E_{\mathrm{s}}$ (see
Eq.(40) from Mewe \& Gronenschild (1981)) corresponding to those
levels. Since the energies $E_{s}$ are smaller than the energy
$E_{\mathrm{M}}$ at which the electron distributions in the Abell
2199 and Coma clusters deviates from a Maxwellian distribution,
the exciting electrons belong to the thermal part of the electron
distribution. Taking into account this fact and the dependence of
the ionic fractions on the distribution function (see Sect.2.1) we
estimate that the decrease of the flux ratio $R_{B}$ for modified
Maxwellian distributions in A2199 and Coma with respect to a
Maxwellian distribution are 30\% and 10\%.

New high spectral resolution instruments with higher sensitivity
such as XEUS are needed to resolve the lines and to measure the
flux ratio of the iron $K_{\alpha}$ lines for the purpose of
testing the hard X-ray tail interpretations.

\begin{acknowledgements}
We are grateful to Jelle Kaastra and Jean-Luc Sauvageot for
valuable discussions and to the referee for constructive comments.

DP and VD are partly supported by the RFBR grant 08-02-00170-a,
the NSC-RFBR Joint Research Project ¹ 95WFA0700088 and by the
grant of the President of the Russian Federation "Scientific
School of Academician V.L.Ginzburg".
\end{acknowledgements}

\end{document}